# Driving a low critical current Josephson junction array with a mode-locked laser

J. Nissilä[1,a], T. Fordell[1], K. Kohopää[1], E. Mykkänen[1], P. Immonen[1], R.N. Jabradaghi[1], E. Bardalen[2], O. Kieler[3], B. Karlsen[4], P.A. Ohlckers[2], R.Behr[3], A.J. Manninen[1], J. Govenius[1], and A. Kemppinen[1]

___

**AFFILIATIONS**

[1]VTT Technical Research Centre of Finland Ltd, 02150 Espoo, Finland

[2]University of South-Eastern Norway, 3184 Borre, Norway

[3]Physikalisch-Technische Bundesanstalt (PTB), 38116 Braunschweig, Germany

[4]Justervesenet, 2007 Keller, Norway

a) Author to whom correspondence should be addressed: Jaani.Nissila@vtt.fi

___

**ABSTRACT**

We demonstrate the operation of Josephson junction arrays (JJA) driven by optical pulses generated by a mode-locked laser and an optical time-division multiplexer. A commercial photodiode converts the optical pulses into electrical ones in liquid helium several cm from the JJA. The performance of our custom-made MLL is sufficient for driving a JJA with low critical current at multiple Shapiro steps. Our optical approach is a potential enabler for fast and energy-efficient pulse drive without expensive high-bandwidth electrical pulse pattern generator, and without high-bandwidth electrical cabling crossing temperature stages. Our measurements and simulations motivate an improved integration of photodiodes and JJAs using, e.g., flip-chip techniques, in order to improve both the understanding and fidelity of pulse-driven Josephson Arbitrary Waveform Synthesizers (JAWS).

___

Josephson junction arrays as voltage standards have been used and investigated for decades [1-4]. The most powerful concept of generating accurate arbitrary voltage waveforms is driving the junctions with ultrafast current pulses [5]. Conventionally, this is done using electrical pulse pattern generators (PPG) and coaxial transmission lines between room temperature (RT) and cryogenic temperature. However, when aiming at higher voltages or frequencies using ever higher pulse rates, this method encounters a limitation because of the limited bandwidth of metallic cables. To overcome this bottleneck and various noise and thermal loading issues, a promising possibility is to use optical pulse patterns and ultrafast photodiodes (PD) at 4 K to drive JJAs [6,7]. A similar technique has recently been used for the control and readout of superconducting qubits [8]. Robust mounting techniques for the PDs at 4 K [9,10] enabled the demonstration of quantized voltage waveforms using pulse rates up to several GHz [11,12]. All these experiments have relied on controlling the intensity of continuous-wave (CW) light with optical intensity modulators (OIM) and expensive high-frequency electronics.

In this letter, we report proof-of-concept experiments on operating a JJA by a cryogenic PD driven with optical pulses generated by a mode-locked laser (MLL) and an optical time-division multiplexer (TDM). MLLs can generate picosecond pulses, which is challenging using continuous-wave light modulated with OIMs driven by electronic PPGs. Our vision is to develop an optical PPG which could be used, e.g., together with ultrafast uni-travelling-carrier (UTC) PDs [13,14] to increase the maximum pulse frequency that drives JJAs. The MLL can be designed to produce ultrashort pulses with very little amplitude variation and time jitter [15,16], which enables low-noise drive signals. In this work, we drive JJAs with an optical PPG and fast commercial PDs [17]. We study the JJA response to current pulses and dc current to develop understanding of both single junction dynamics and the system as a whole. All in all, we demonstrate that the performance of our custom-made MLL is sufficient for driving a JJA with low critical current ($I_c$) at multiple Shapiro steps.

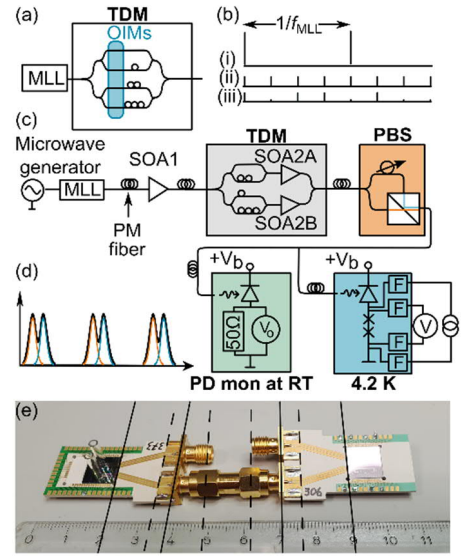

*Fig. 1. (a) Simplified illustration of an optical pulse pattern generator with a mode-locked laser (MLL), time-division multiplexer (TDM) and optical intensity modulators (OIM) for pulse picking. (b) An example of pulse pattern at (i) the MLL, (ii) the TDM output when modulators pass all pulses and (iii) the TDM output when some pulses are blocked. (c) A more detailed diagram of the PPG used in this paper. SOAxx = semiconducting optical amplifiers, PD mon = monitoring photodiode, PBS = polarization based splitter, F = low-pass filters. (d) Illustration of two pulses with orthogonal polarizations (blue and red) summed by the PBS to form a composite pulse, (e) Photograph of the cryogenic assembly with PD on the left with a glass tube for fiber mounting, and JJA on the right. The solid and dashed lines show the possible non-ideal mismatched interfaces in the high-speed transmission line of this preliminary experimental setup. Two similar PD-JJA channels are visible, but only the one in front is connected.*

A Josephson junction array quantizes electric pulse patterns such that the voltage across the array equals $V = Ns(t) \times \frac{h}{2e} f(t)$ [2-4]. Here $N$ is the number of junctions in series, $s$ is an integer describing the Shapiro step index to which the array is biased at time $t$, and $f(t)$ is the time-dependent pulse frequency. The ratio $h/2e$, where $h$ is the Planck constant and $e$ is the elementary charge, is equal to the magnetic flux quantum. When aiming at increased voltages, an evident possibility is to use higher pulse rates or higher Shapiro steps, or both. The latter is easier for JJAs with low $I_c$ since the linearity, dissipation and distortion issues in both the photodiode and the JJA suggest minimizing the current pulse integrals. However, if $I_c$ is too low, the shot noise of PD and Josephson coupling energy may become limitations. In this study, we focus on investigating our optical pulse source driving a JJA with $I_c = 360\ \mu A$, which is considerably lower than typically used values in metrology.

Figure 1 (a) shows the basic structure of our optical PPG. The MLL produces fast pulses in the wavelength range from 1200 nm to 1700 nm at a modest repetition rate of 1-10 GHz. The TDM splits the parent pulses into several waveguides without introducing additional noise. Each path has an OIM which is used either to pass or block pulses. Modulator state transitions are arranged to happen simultaneously and well before the arrival of pulses, enabling negligible distortion to the passing pulses. These pulses are interleaved with proper delays and combined back to a single waveguide. Fig. 1 (b) illustrates an example of pulse pattern evolution in a system which enables multiplication of the MLL pulse repetition rate $f_{MLL}$ by factor 4.

Figure 1 (c) illustrates our proof-of-concept experiment. We have designed and built an inexpensive MLL at 1335 nm wavelength which is suitable for generating charge carriers in regular InGaAs photodiode at 4.2 K. The pulse repetition rate is adjustable in the range of $f_{MLL} = 2\ldots2.5$ GHz and it was set to $f_{MLL} = 2.3$ GHz in this experiment. The laser cavity is ring-type where the pulsing is excited using both a semiconducting saturable absorber mirror (SESAM) and by varying the gain of a semiconducting optical amplifier (SOA) driven by a microwave generator. The transform-limited full width at half maximum (FWHM) of our laser pulses is 6 ps, but the photodiode broadens the pulses to approximately 25 ps in this experiment. The MLL pulses are further amplified with an SOA and guided into the TDM, which multiplies the maximum pulse frequency by factor 4. Our TDM is composed of fiber-optic splitters and four adjustable delay lines to interleave outcoming pulses. SOAs inside the multiplexer are used to increase the optical power to a sufficient level for the JJAs.

Our optical PPG does not yet have optical intensity modulators, but we can test JAWS-like operation by blocking some of the optical paths. In addition, we have a polarization-based splitter (PBS) that can increase the maximum pulse frequency by a factor of 2, hence enabling the maximum pulse frequency of $2 \times 4 \times 2.3$ GHz = 18.4 GHz. The PBS splits the incoming linearly polarized pulses into two beams with orthogonal polarizations before combining them again. Both time interval $\Delta t$ and amplitudes of the pulses can be set to any values without distorting the shape of individual pulses. The pulses can even be set overlapping without interference induced instabilities.

The optical pulses were brought to the PD with a polarization maintaining (PM) single-mode fiber. It is terminated with a glass ferrule, which is mounted in a glass tube such that the optical alignment is stable over the experimental time. A commercial bare chip InGaAs p-i-n photodiode with 28 GHz bandwidth [17] was flip-chip bonded on a separate silicon carrier with a 50-ohm Nb coplanar waveguide (CPW) with 120 pF finger capacitors in the ground lines for fast release of charge during the pulses. The PD carrier and the JJA were mounted on separate printed circuits boards, which were connected together with an SMA connector (Fig. 1 (e)). In this way, the electrical connection between the PD and JJA is direct with no dc-blocking capacitors as often used in conventional pulse-driven JAWS. In addition, we split the optical pulses also to a similar PD at room temperature in order to approximately monitor the output pulses with an oscilloscope.

We have measured two Superconductor-Normal metal-Superconductor (SNS) -type JJAs fabricated at PTB. They have $N = 1000$ Nb-$Nb_xSi_{1-x}$-Nb Josephson junctions in series along the center conductor of a wide-band superconducting CPW [18,19]. Array A has $I_c$ = 360 µA and junction resistance $R_J$ = 30.8 mΩ, which yield the characteristic frequency [5] $f_c$ = 5.4 GHz. Array B has $I_c$ = 2.3 mA, $R_J$ = 11.4 mΩ, and $f_c$ = 12.7 GHz. To compensate the dissipation of the JJA the insulating gap of the CPW narrows down linearly so that the transmission line impedance decreases from 50 Ω to 35 Ω [20]. The ideal impedance at the end of the array would be 50 Ω − $NR_J$, i.e. 19 Ω and 39 Ω for arrays A and B, respectively. More details on the experimental setup can be found in Suppl. A and [9,10,12].

In the experiments, we have used TDM and PBS (Fig. 1 (c)) to generate pulse pairs with a variable $\Delta t$ and repetition rate 2.3 GHz. The amplitude of the pulse pair was tuned by changing the amplification of SOA1.

In Fig. 2 we present experimental and simulated average voltage of the JJAs for array A with five different values of $\Delta t$, varying from 30 ps to 220 ps (see panel (g)). Panels (a) - (f) show the results for normalized voltage $v = V/(N \times \frac{h}{2e} \times f_{MLL})$ as a function of normalized pulse integral $p$ and supplementary dc bias current $I_{dc}$. We define $p = \frac{2\pi f_c}{I_c} \int_0^{1/f_{MLL}} I_p \, dt$, where $I_p = I - I_{dc}$ is the pulse current. At the quantized voltage plateaus, $v$ is ideally an integer $v = s \times m$, where $m$ is the number of pulses in the period $1/f_{MLL}$ (in our case $m = 2$). The simulations of panels (b), (d), and (f) were performed with a single junction model (Resistively and capacitively shunted Josephson junction model, RCSJ) that omits any transmission line effects [21,22] (see Suppl. B.1).

The characteristic time for array A is $t_c = 1/f_c$ = 190 ps. For $\Delta t > t_c$, the JJA has sufficient time to recover after each pulse, and we see quantized plateaus only with even $v = 0, 2, 4, …$ In the opposite limit, $\Delta t \ll t_c$, the JJA sees the two pulses as a single one with $p = p_1 + p_2$, where $p_i$ are the integrals of the individual current pulses. In this limit, we expect to see all plateaus $v = 1, 2, 3, …$ For intermediate $\Delta t$ we also expect to see all plateaus, but plateaus with even index should be wider than those with odd index. The results of our measurements and single-junction simulations are in qualitative agreement.

In the following discussion we concentrate on the important $I_{dc} = 0$ data shown in Fig.2(h). We have scaled the pulse integral data measured from the room temperature monitor PD (Fig. 2 (g)) with a single fitting parameter. It was chosen such that the transition from $v = 0$ to $v = 1$ with $\Delta t = 30$ ps occurs at the same pulse integral when $I_{dc} = 0$ (see Fig. 2 (h)).

The single-JJ simulations predict that the width of even plateaus increases with increasing $\Delta t$, but this effect is significantly weaker in the measured data. Especially, the experimentally measured plateau $v = $

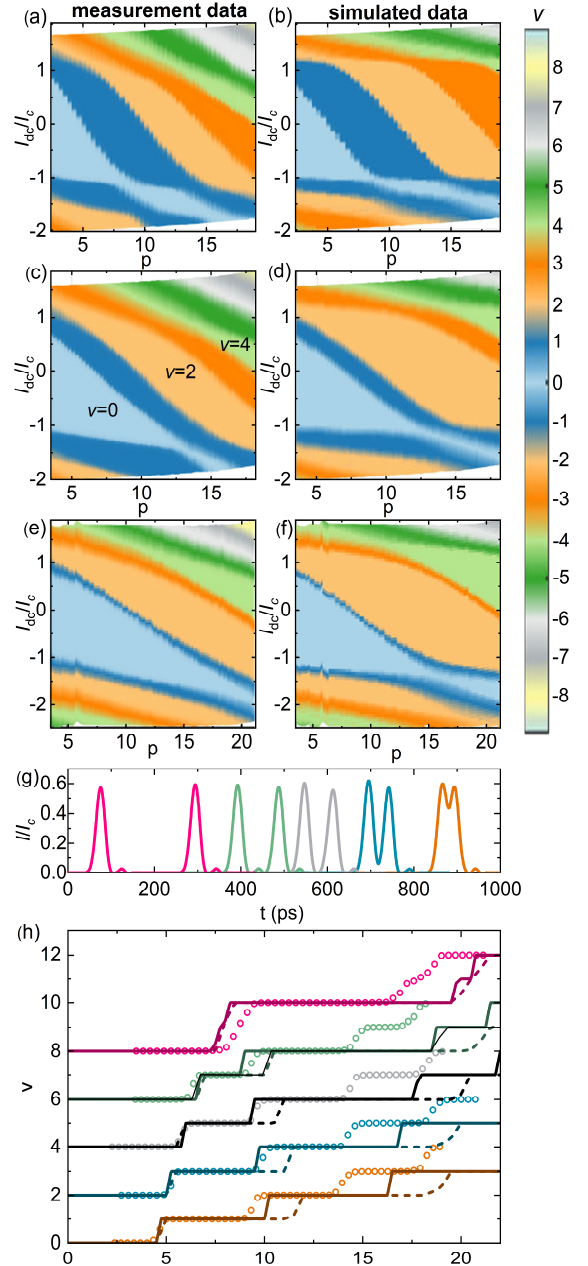

*Fig. 2. (a)-(e) Measured and simulated average JJA voltage as a function of normalized current pulse time-integral and normalized dc current through the array. Current pulse pairs with repetition rate 2.3 GHz (period 430 ps) with time intervals $\Delta t$ of 30 ps ((a) and (b)), 95 ps ((c) and (d)), and 220 ps ((e) and (f)) were applied. (g) Current pulses composed of two elementary pulses set at different time intervals from each other: 30 ps (orange), 45 ps (blue), 70 ps (grey), 95 ps (green), and 220 ps (red). These pulse patterns are based on measurements using a monitoring photodiode at room temperature. (h) Comparison of measured and simulated average voltages at zero dc current as a function of normalized pulse integral. The circles depict the experimental values, the solid lines show the simulated results using a single-JJ model, and the dashed lines the simulated values using a transmission line model for the JJA (1000 JJ) and an unterminated photodiode in 10-mm-long 50 Ω transmission line. The thin black line describes the calculated voltage using the distorted pulse train (Suppl. B.3). Different traces correspond to different $\Delta t$ with the same colors as in (g), and they have been shifted vertically for clarity.*

2 (corresponding to $V \approx 9.5$ mV) has nearly equal width for all values from $\Delta t = 30$ ps to $\Delta t = 95$ ps. However, thanks to the adjustability of the $\Delta t$ between the short pulses, we can observe experimentally the crossover at $\Delta t = t_c$: as expected, odd steps are missing with the longest $\Delta t = 220$ ps $> t_c$, but they are observable when $\Delta t < t_c$.

Qualitatively similar observations, i.e. that plateaus with odd $v$ were wider than expected for $\Delta t \approx 100$ ps, were also obtained for array B (data not shown). Since it had a notably shorter characteristic time $t_c = 80$ ps, it is clear that the results cannot be explained using our simple model. Our model omits the heating of the JJ, which is essentially proportional to $R_J I_{ave}^2$ where $I_{ave}^2$ is the time average of the total current squared through the junction. Heating is thus much more significant in array B that has a larger $I_c$. We cannot completely overrule heating effects at the highest Shapiro steps and dc currents, but the qualitative similarity of results between arrays A and B also indicates that heating does not explain the discrepancy between experiments and 1-JJ simulations.

Since the single-JJ model calculations do not explain the experimental observations quantitatively, we have built a transmission line model for the JJA (see Suppl. B.2). The PD is modelled as an ideal current source and the end of the JJA is terminated with a 35 Ω resistor. We first modeled only the JJA, i.e., the PD was in the vicinity of the first junction of the JJA. In this case, the single-JJ and transmission line models agree with minor differences.

Figure 3 demonstrates the effect of adding a passive transmission line between PD and JJA corresponding to the more realistic case of a non-negligible distance between the elements. Instead of modeling details of the highly nonideal transmission line shown in Fig. 1 (e), we made simulations of the effect of 10 mm of ideal 50 Ω transmission line between the PD and the JJA. Such a distance may be realistically required to avoid overheating of the JJA if gold-based flip-chip bonding is used for mounting the PD. Any non-ideality of the JJA transmission line will reflect part of the drive pulses back to the PD. Since the PD has a high impedance, most of the energy reflects back to the JJA.

In Fig. 3, the simulation results correspond to a steady state situation at the first and last JJ of the arrays A and B. The reflections are stronger in array A, which we relate to the less ideal impedance matching of $R_J$. We notice both rapid variation in the background with time as well as period-long fluctuation. The latter resembles to non-intentionally varied dc current background. In panel (c) the amplitude of background current fluctuation is of the order of $I_c$. Considering the voltage maps in Fig. 2, this large fluctuation can cause robustness issues in voltage standard use. In our simulations, we have observed events where the preceding pulse pattern affects the JJA response e.g. such that different number of flux quanta are transferred through the JJs along the array. Data for array A are shown at the edge of a quantized voltage plateau where the current pulse reflections have the strongest effect on the resulting voltage. However, in the middle of the step the background fluctuations in current are of the same order of magnitude (data not shown). The simulation data for array B are shown in the middle of the plateau $v = 2$, where the voltage is insensitive to differences of current pulses. Since JJs are nonlinear components, we expect that the JJA will cause reflections also for more ideally matched transmission lines especially with faster pulses. Furthermore, if multiple Shapiro steps are used in voltage generation, the background easily becomes code dependent which impairs the robustness in voltage standard usage.

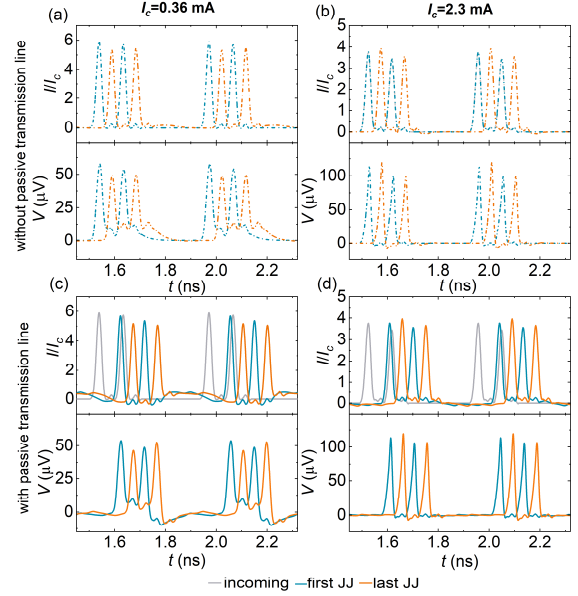

Fig. 3. Simulated current and voltage pulse waveforms at the first and last JJ in a JJA for arrays A and B with two pulses separated by 95 ps. Panels (a)-(b) show the simulated current and voltage for the case without any transmission line between PD and JJA for arrays A and B, respectively. Panels (c)-(d) show the corresponding simulations when there is a 10 mm transmission line between PD and JJA. The simulations shown here for array A (panels (a),(c)) were performed close to the beginning of plateau $v = 2$ whereas simulations for array B ((b),(d)) were performed in the middle of plateau $v = 2$.

As shown by the dashed lines in Fig. 2(h), the transmission line has a big effect on the position of the quantized current plateaus. However, these simulations do not explain the observed difference to the experimental data, and actually the discrepancy is larger than with the single-JJ model.

In the analysis of data in Fig. 2 we applied a scaling factor to match the measured transition from $v = 0$ to $v = 1$ to the simulated data. This factor is 28 % smaller than expected from measurements of the total pulse current through the PD and JJA. This suggests that nearly 30 % of the pulse amplitude may be lost in reflections in the PD-JJA assembly. We tested this hypothesis by a simulation in which the incoming current pulse train is distorted in such a way that 80% of current is in the main pulse pair and 20% in another pulse pair delayed by 50 ps (see Suppl. B.3). The effect on the $v(p)$ curve for $\Delta t = 95$ ps is that the beginning of the $v = 3$ plateau moves towards lower pulse integral which agrees better with the experimental observation.

To conclude, we have demonstrated an optically driven Josephson voltage standard based on generating fast optical pulses with a mode-locked laser. Proper quantization of current pulses into quantized voltages takes place using a JJA with a relatively low critical current $I_c = 360$ µA. Sharp transitions between Shapiro steps indicate that the optical pulses have sufficiently low noise for metrological operation of the JJA. Our results pave the way for using higher Shapiro steps and higher pulse frequencies to obtain higher output voltages in Josephson Arbitrary Waveform Synthesizers. However, comparison between measurements and simulations shows that in the future, the PD should be mounted close to a JJA, and the PD should possibly be coupled using more complex circuitry in order to provide an impedance matched source.

**ACKNOWLEDGEMENT**

We are grateful to Dr. Mark Bieler from PTB for discussions. This work was partly funded by the Academy of Finland grant 310909. TF received funding from the Academy of Finland (grants 296476 and 306844). The work is also part of the Academy of Finland Flagship Programme, Photonics Research and Innovation (PREIN), decision 320168. Part of the work has been carried out in a Joint research project QuADC which has received funding from the EMPIR program co-financed by the Participating States and from the European Union's Horizon 2020 research and innovation program.

**REFERENCES**

[1] B.D. Josephson, Phys. Lett. **1**, 251 (1962).
[2] C.A. Hamilton, Rev. Sci. Instrum. **71**, 3611 (2000).
[3] J. Kohlmann, R. Behr, T. Funk, Meas. Sci. Technol. **14**, 1216 (2003).
[4] B. Jeanneret and S.P. Benz, Eur. Phys. J. Special Topics **172**, 181 (2009).
[5] S. P. Benz and C. A. Hamilton, Appl. Phys. Lett. **68**, 3171 (1996).
[6] C. Urano, N. Kaneko, M. Maezawa, T. Itatani, and S. Kiryu, J. Phys.: Conf. Ser. **97**, 012269 (2008).
[7] J. Williams, T. J. B. M. Janssen, L. Palafox, D. A. Humphreys, R.Behr, J. Kohlmann, and F. Müller, Supercond Sci.Technol. **17**, 815 (2004).
[8] F. Lecocq, F. Quinlan, K. Cicak, J. Aumentado, S. A. Diddams, J. D. Teufel, arXiv:2009.01167 [quant-ph] (2020).
[9] E. Bardalen, B. Karlsen, H. Malmbekk, O. Kieler, M. N. Akram, and P. Ohlckers, IEEE Trans.Comp. Pack.Manuf. Tech. **7**, 1395 (2017).
[10] E. Bardalen, B. Karlsen, H. Malmbekk, O. Kieler, M. N. Akram, and P. Ohlckers, Microelectronic Reliab. **81**, 362 (2018).
[11] O. Kieler, B. Karlsen, P.A. Ohlckers, E. Bardalen, M.N. Akram,R. Behr, J. Ireland, J. Williams, H. Malmbekk, L. Palafox and R. Wendisch, IEEE Trans. Appl. Supercond. **29**, 1200205 (2019).
[12] B. Karlsen, O. Kieler, R. Behr. T.A.T Nguyen, H. Malmbekk,M. N. Akram, and P. Ohlckers, IEEE Trans. Appl. Supercond. **29,** 1200308 (2019).
[13] T. Ishibashi, N. Shimizu, S. Kodama, H. Ito, T. Nagatsuma and T. Furuma, Tech.Dig. Ultrafast Electronics and Optoelectronics, 166 (1997).
[14] T. Kurokawa, T. Ishibashi, M. Shimizu, K. Kato, and T. Nagatsuma, Electron. Lett. **54**, 705 (2018).
[15] F. Quinlan, S. Gee, S. Ozharar, and P. J. Delfyett, Optics Letters **31**, 2870 (2006).
[16] D. Kim, D. Kwon, B. Lee, and J. Kim, Opt. Lett. **44**, 1068 (2019).
[17] Albis Optoelectronics, Switzerland, Data Sheet for PD20X1 28 Gb/s photodiode with integrated lens.
Available: www.albisopto.com/albisproduct/pd20x1/
[18] B. Baek, P. D. Dresselhaus, and S. P. Benz, IEEE Trans. Appl. Supercond. **16**, 1966 (2006).
[19] O. Kieler, R. Behr, D. Schleussner, L. Palafox, and J. Kohlmann, IEEE Trans. Appl. Supercond. **23**, 1301404 (2013).
[20] P.D. Dresselhaus, M.M. Elsbury and S.P. Benz, IEEE Trans. Appl. Supercond. 19, 993 (2009).
[21] W.C.Stewart, Appl. Phys. Lett. **12**, 277 (1968)
[22] D.E.McCumber, Journal of Applied Physics **39**, 2503 (1968).

# *Supplementary information*

**A. Experiments**

The arrays were mounted in a cryoprobe enabling immersion in liquid helium bath. Copper wires were used for measurement of voltage across the JJA and stainless steel coaxial cable for driving dc current through the array. The JJA was fabricated on a silicon carrier chip with Nb coplanar waveguide (CPW) wire-bonded to a pcb with a CPW ending at an SMA connector.

The photodiode driving the JJA was biased from a linear voltage source at 5 V through a current meter and 0.1-mm-diameter 1.5-mm-long copper wires in the cryoprobe. In order to verify that the voltage source does not introduce marked noise in the system, we have also tested biasing the PD from a battery with no difference observed. The voltage signal across the JJA was measured with a commercial nV meter using 100 mV scale and usually 100 ms integration time. The dc bias current through the array was supplied by a low-noise current source. The monitor photodiode pulses were measured using a 26.5 GHz bandwidth sampling oscilloscope through a -20 dB SMA attenuator. The optical pulse height was controlled by varying the injection current of SOA1 at the output of the MLL.

.

## B. Simulations

### B.1. A single junction

Simulating the whole JJA was very time consuming and therefore the 3D plots of Fig. 2 were done by approximating the JJA with a single junction and utilizing the well-known form for resistively shunted junction [21,22]

$$\frac{d\phi}{d\tau} = \frac{I}{I_C} - \sin\phi, \quad (1)$$

where $\phi$ is the phase over the junction and $\tau = 2\pi f_c t$ is normalized time ($t$ is time in SI units). We solve Eq. (1) directly in time domain using discrete time steps: $\phi_j = \phi_{j-1} + d\phi_{j-1}dt$.

### B.2. Multiple junctions in series (array)

In our JJA, JJs are located in the center conductor of a coplanar transmission line. Each junction has the length of 4 μm along the transmission line, and the distance between junctions is 2 μm. We approximate this in our model with an elementary section that consists of a point-like JJ in series with an inductor that corresponds to 6 μm of transmission line. The transmission line inductance and capacitance were different for each section to take into account the change of the transmission line impedance. The PD is modelled as an ideal current source and the end of the JJA is terminated with a 35 Ω resistor.

To simulate the whole JJA (see Fig. 1), we use a custom simulator developed at VTT. The program can solve a circuit with any number of resistors, inductors, capacitors and Josephson junctions, and is based on direct time-domain simulation similar to the simple model (1). However, instead of solving the time dependence for a scalar phase $\phi$, we solve it for a vector $\boldsymbol{f}_j = \boldsymbol{f}_{j-1} + d\boldsymbol{f}_{j-1}dt$, where $\boldsymbol{f} = [\boldsymbol{\phi}, \boldsymbol{V}]$, and $\boldsymbol{\phi}$ and $\boldsymbol{V}$ refer to phase and voltage vectors, respectively. The time derivative of phase is obtained from $\frac{d\phi}{dt} = \frac{2e}{h}V$.

The simulation requires the circuit to be divided to discrete impedances between nodes (see Figure 2). The derivative of voltage, $d\boldsymbol{V}_{j-1}$, is calculated by inverting the capacitance matrix of the circuit and multiplying the circuit currents with it, i.e. $\boldsymbol{C}^{-1}\boldsymbol{I}$. Currents depend on the type of individual impedances. For resistors and Josephson junctions these can be written as $I_R = R^{-1}V$, and $I_{JJ} = I_c \sin\phi$, where $R$ is resistance and $I_c$ is critical current of a Josephson junction. For inductors one needs to calculate the inductance matrix $\boldsymbol{L}$ of the subset of nodes connected by coupled inductors. Then the current vector of inductive elements can be calculated as $\boldsymbol{I}_L = \boldsymbol{L}^{-1}\Delta\boldsymbol{\phi}$, where elements in $\Delta\boldsymbol{\phi}$ are the difference of phase differences between nodes.

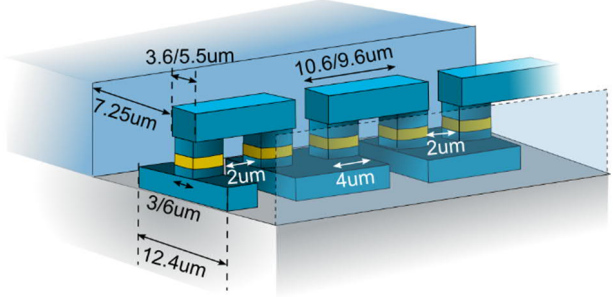

*Figure 1. Illustration of the JJA. The yellow parts are the normal-metal elements of the JJs. The lower plane defines the CPW.*

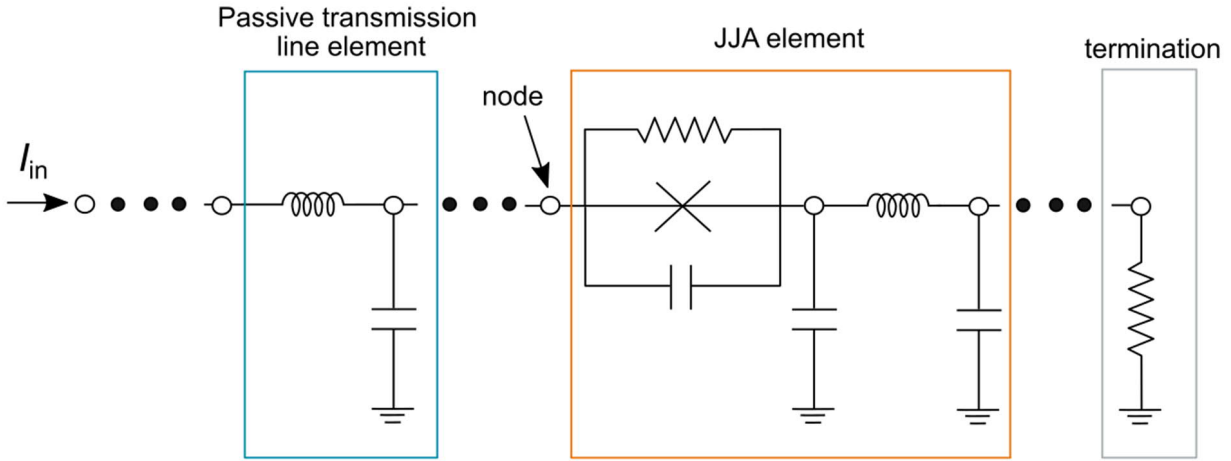

*Figure 2: Schematics of the JJA architecture in the simulation program. The circuit is divided to impedances, Z, between nodes. There can be any number of impedances in parallel between two nodes and any number of nodes. The circuit is terminated to a ground node. An impedance can consist of one of the following: capacitance, inductance, resistance or Josephson junction. If one wants to add two of these components in series, one needs to put a node in between.*

## B.3. Distorted current pulse train

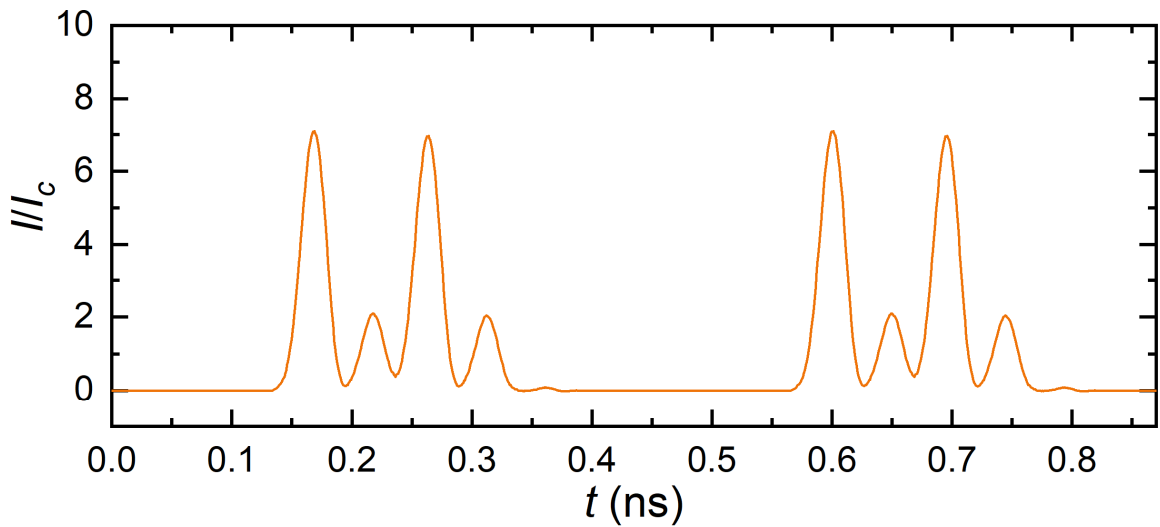

*Fig. 3. Current pulse train which was used in simulations to analyze the effect of a markedly distorted pulse pair entering the JJA. A pulse pair with dt=95 ps has been splitted in two parts, with 80 % amplitude in the main part and 20 % fraction following after a 50 ps delay.*